\documentclass[twocolumn,aps]{revtex4}
\usepackage{graphics}
\begin{document}
\title{\LARGE Persistent currents with long-range hopping in $1$D  
single-isolated-diffusive rings}
\author{{\Large Santanu K. Maiti} \footnote{Corresponding author: 
Santanu K. Maiti \\
E-mail: santanu@cmp.saha.ernet.in}}
\author{\Large J. Chowdhury}
\author{\Large S. N. Karmakar}
\affiliation{\large Saha Institute of Nuclear Physics, 1/AF, Bidhannagar,
Kolkata 700 064, India}
\begin{abstract}
\noindent
{\bf\normalsize Abstract}
\vskip 0.1cm
{\normalsize We show from exact calculations that a simple tight-binding Hamiltonian
with diagonal disorder and long-range hopping integrals, falling off as
a power $\mu$ of the inter-site separation, correctly describes the 
experimentally observed amplitude (close to the value of an ordered ring) 
and flux-periodicity ($hc/e$) of persistent currents in 
single-isolated-diffusive normal metal rings of mesoscopic size. 
Long-range hopping integrals tend to delocalize the electrons even in the 
presence of disorder resulting orders of magnitude enhancement of persistent
current relative to earlier predictions.}
\end{abstract}
\maketitle

The phenomenon of persistent current in mesoscopic normal metal rings has 
generated a lot of excitement as well as controversy over the past years. 
In a pioneering work, B\"{u}ttiker, Imry and Landauer~\cite{butt} predicted 
that, even in the presence of disorder, an isolated $1$D metallic ring 
threaded by magnetic flux $\phi$ can support an equilibrium persistent 
current with periodicity $\phi_0=ch/e$, the flux quantum. Later, experimental 
observations confirm the existence of persistent currents in isolated 
mesoscopic rings. However, these experiments yield many results that are 
not well-understood theoretically even today~\cite{cheu1,cheu2,mont,bouc,
alts,von,schm,ambe,abra,bouz,giam,burme}. The results of the single loop 
experiments are significantly different from those for the ensemble of 
isolated loops. Persistent currents with expected $\phi_0$ periodicity 
have been observed in isolated single Au rings~\cite{chand} and in a 
GaAs-AlGaAs ring~\cite{maily}. Levy {\em et al.}~\cite{levy} found 
oscillations with period $\phi_0/2$ rather than $\phi_0$ in an ensemble 
of $10^7$ independent Cu rings. Similar $\phi_0/2$ oscillations were 
also reported for an ensemble of disconnected $10^5$ Ag rings~\cite{deb} 
as well as for an array of $10^5$ isolated GaAs-AlGaAs rings~\cite{reul}. 
In a recent experiment, Jariwala {\em et al.}~\cite{jari} obtained both 
$\phi_0$ and $\phi_0/2$ periodic persistent currents in an array of thirty 
diffusive mesoscopic Au rings. Except for the case of the nearly ballistic 
GaAs-AlGaAs ring~\cite{maily}, all the measured currents are in general 
one or two orders of magnitude larger than those expected from the 
theory~\cite{cheu1,cheu2,bouc,mont,von,alts,schm,ambe,abra,bouz,giam}. 
The diamagnetic response of the measured $\phi_0/2$ oscillations of 
ensemble-averaged persistent currents near zero magnetic field also 
contrasts with most predictions~\cite{schm,ambe}. 

 Free electron theory predicts that at $T=0$, an ordered $1$D metallic 
ring threaded by magnetic flux $\phi$ supports persistent current with 
maximum amplitude $I_0=ev_F/L$, where $v_F$ is the Fermi velocity and $L$ 
is the circumference of the ring. Metals are intrinsically disordered 
which tends to decrease the persistent current, and the calculations 
show that the disorder-averaged current $<I>$ crucially depends on the 
choice of the ensemble~\cite{cheu2,mont,bouc}. The magnitude of the current 
$<I^2>^{1/2}$ is however insensitive to the averaging issues, and is of 
the order of $I_0l/L$, $l$ being the elastic mean free path of the electrons. 
This expression remains valid even if one takes into account the finite 
width of the ring by adding contributions from the transverse channels, 
since disorder leads to a compensation between the channels~\cite{cheu2,mont}. 
However, the measurements on an ensemble of $10^7$ Cu rings~\cite{levy} 
reported a diamagnetic persistent current of average amplitude $3\times 
10^{-3} ev_F/L$ with half a flux-quantum periodicity. Such $\phi_0/2$
oscillations with diamagnetic response were also found in other
persistent-current experiments consisting of ensemble of isolated
rings~\cite{deb,reul}. 

Measurements on single isolated mesoscopic rings on the other hand detected 
$\phi_0$-periodic persistent currents with amplitudes of the order of 
$I_0\sim ev_F/L$, (closed to the value for an ordered ring). Theory and 
experiment~\cite{maily} seem to agree only when disorder is weak. However, 
the amplitudes of the currents in single-isolated-diffusive gold 
rings~\cite{chand} were two orders of magnitude larger than the theoretical 
estimates. This discrepancy initiated intense theoretical activity, and it 
is generally believed that the electron-electron correlation plays an 
important role in the disordered diffusive rings~\cite{abra,bouz,giam}, 
though the physical origin behind this enhancement of persistent current 
is still unclear. 

 In this letter we will address the problem of enhancement of persistent
current in single-isolated-diffusive (SID) mesoscopic rings. The large 
amplitude of the observed currents in SID mesoscopic rings strongly 
challenges the conventional theories of persistent current. It indicates 
that in all the previous models some fundamental mechanism is missing 
which could compensate the effect of impurities, and thus prevents 
reduction of the current due to disorder. The existing theories are 
basically within the framework of the Anderson model where the transport 
properties of the electrons are dominated by the localization phenomenon 
that essentially reduces the persistent current. In a recent work, 
Balagurov {\em et al.}~\cite{bala} have shown that the electrons become 
delocalized if one includes long-range hopping (LRH) integrals in the 
Anderson model. 

 We describe a $N$-site ring enclosing a magnetic flux $\phi$ (in units 
of the elementary flux quantum $\phi_0$) by the following Hamiltonian in 
the Wannier basis
\begin{equation}
H=\sum_i\epsilon_i c_{i}^{\dagger}c_{i}
+\sum_{i\ne j} v_{i,j}\left[e^{-i \theta} c_{i}^
{\dagger} c_{j}+ h.c. \right]
\label{hamil}
\end{equation} 
where $\epsilon_i$'s are the site energies, $v_{i,j}$'s are the hopping
integrals, and $\theta=\frac{2\pi\phi}{N}(|i-j|)$. The $\epsilon_i$'s are 
uncorrelated random variables drawn from some distributive function 
$P(\epsilon)$, and, the non-random LRH integrals are taken as $v_{i,j}=
v/|i-j|^{\mu}$, $v$ being a constant representing the nearest-neighbor 
hopping (NNH) integrals. It should be noted that the physical domain of 
the exponent $\mu$ is determined by the boundness of the spectrum~\cite{bala}. 
The choice of the distribution function $P(\epsilon)$ which we shall use 
are the ``box" distribution
\begin{equation}
P(\epsilon)=\frac{\Theta\left(\frac{W}{2}-|\epsilon|\right)}{W} 
\label{site1}
\end{equation}
of width $W$, and the ``binary-alloy" distribution
\begin{equation}
P(\epsilon)=c\delta\left(\epsilon-\epsilon_A\right)+(1-c)\delta
\left(\epsilon-\epsilon_B\right)
\label{site2}
\end{equation}
where $c$ and $(1-c)$ are respectively the concentrations of two types of 
atoms with site energies $\epsilon_A$ and $\epsilon_B$. In the following 
we use the units $h=e=1$.

 For an ordered ring, setting $\epsilon_i=0$ for all $i$, the energy
of the $n$th eigenstate can be expressed as
\begin{equation}
E_{n}(\phi)=\sum_{m=1}^{N-1}\frac{2v}{m^{\mu}}\cos\left[\frac{2\pi m}{N}
\left(n+\phi \right)\right]
\label{energy}
\end{equation}
where $m$ is an integer. The current carried by this eigenstate is 
given by
\begin{equation}
I_{n}(\phi)=\left(\frac{4\pi v}{N}\right)\sum_{m=1}^{N-1}m^{(1-\mu)}
\sin\left[\frac{2\pi m}{N}\left(n+\phi \right)\right].
\label{current}
\end{equation}

\begin{figure}[ht]
{\centering \resizebox*{8.5cm}{9.5cm}{\includegraphics{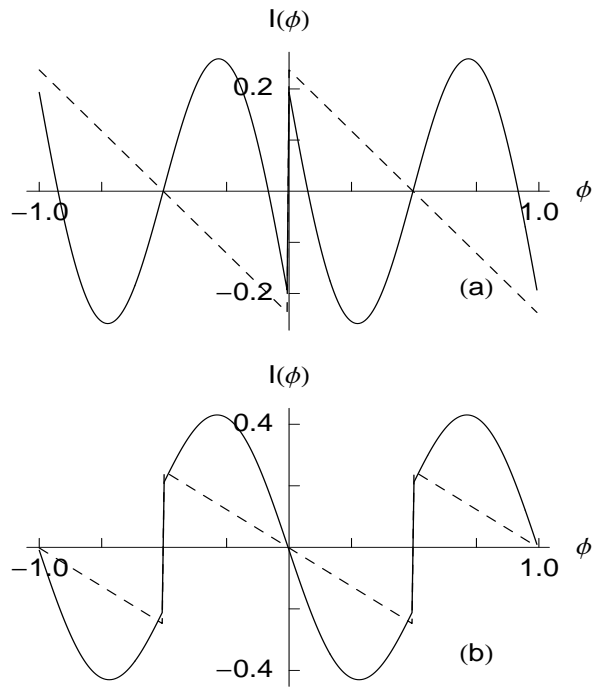}}\par}
\caption{$I-\phi$ characteristics of ordered spinless fermionic rings with
$\mu=1.6,~ v=-1,~ N=50$, and, a) $N_e=20$ and b) $N_e=23$. The solid and
dashed lines are respectively for the rings with all LRH and only NNH
integrals.}
\label{spinless}
\end{figure}
For spinless electrons, we can express the total current at $T=0$ in the 
following form
\begin{equation}
I(\phi)=\sum_{n} I_n(\phi)
\label{spinlesscurrent}
\end{equation}
where $N_e$ is the number of electrons and 
$-\lfloor N_e/2\rfloor \leq n < \lfloor N_e/2 \rfloor$ ($\lfloor z\rfloor$ 
denotes the integer part of $z$). In the above expression we restrict 
$\phi$ in the domains $-0.5 \leq \phi <0.5$
and $0 \leq \phi < 1$ for the systems respectively with odd and even number 
of electrons. The $I(\phi)$ versus $\phi$ curves for some representative
impurity-free systems are plotted in Fig.~\ref{spinless}. The persistent 
current as a function of $\phi$ always exhibits discontinuity at certain 
points as long as there is no impurity in the system. The ground states 
are degenerate at these points of discontinuity due to the crossing of 
the energy levels. In the presence of all LRH integrals it is found that 
the amplitude of the current initially increases as we increase the system 
size, but eventually it falls when the system becomes larger. This is due 
to the fact that as we increase the number of sites, the Hamiltonian 
Eq.~(\ref{hamil}) includes some additional higher order hopping integrals 
which causes an increase in the net velocity of the electrons, but after 
certain system size this increment in velocity drops to zero because the 
additional hopping integrals are then between far enough sites giving 
negligible contributions.

 If we take into account the spin of the electrons, then the total 
persistent current in an ordered ring with even number of electrons 
is given by
\begin{equation}
I(\phi)=2\sum_{n} I_n(\phi)
\label{fermioniceven}
\end{equation}
where $-\lfloor N_e/4 \rfloor \leq n < \lfloor N_e/4 \rfloor$, and here we
restrict $\phi$ in the domain $0 \leq \phi < 1$ if $N_e/2$ is even while 
in the domain $-0.5 \leq \phi < 0.5$ if $N_e/2$ is odd. If the system 
contains odd number of electrons, we have
\begin{equation}
I(\phi)=2\sum_{n} I_n(\phi) + I_{n^{\prime}}(\phi)
\label{fermionicodd}
\end{equation}
where $-\lfloor \left(N_e-1\right)/4 \rfloor \leq n < \lfloor 
\left(N_e-1\right)/4 \rfloor$.
The quantum number $n^{\prime}$ has to be determined in the following way. 
\begin{figure}[ht]
{\centering \resizebox*{8.5cm}{9.5cm}{\includegraphics{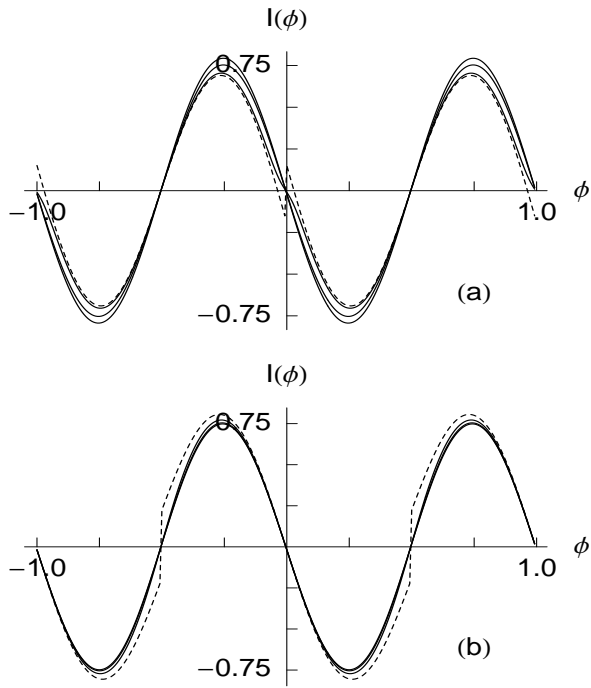}}\par}
\caption{$I-\phi$ curves of spinless fermionic rings with all LRH integrals. 
$\epsilon_i$'s are chosen randomly from Eq.~(\ref{site1}) and the parameters are
$\mu=1.4,~v=-1,~N=50$, and, a) $N_e=20$ and b) $N_e=23$. The dotted curve 
and the three solid curves in each of these figures are respectively
for the ordered and three microscopic disordered configurations
of the ring. }
\label{random}
\end{figure}
Any odd value of $N_e$ can be expressed into the form $(4p\pm 1)$ where 
$p=1,2,3,\ldots$, and the quantum number $n^{\prime}$ becomes equal to 
$\pm p$ corresponding to these two forms of $N_e$. In the above expression 
we restrict $\phi$ in the range $0 \leq \phi < 1$ when $N_e$ is of the form 
$(4p-1)$, whereas $\phi$ is to be bounded between $-0.5 \leq \phi < 0.5$ when
$N_e$ has the form $(4p+1)$. We do not display the $I-\phi$ characteristics 
for spin fermionic systems as in the absence of electron-electron interaction, 
the electron spin cannot alter the characteristic features of the persistent 
currents from those presented in Fig.~\ref{spinless}. 

\begin{figure}[ht]
{\centering \resizebox*{8.5cm}{9.5cm}{\includegraphics{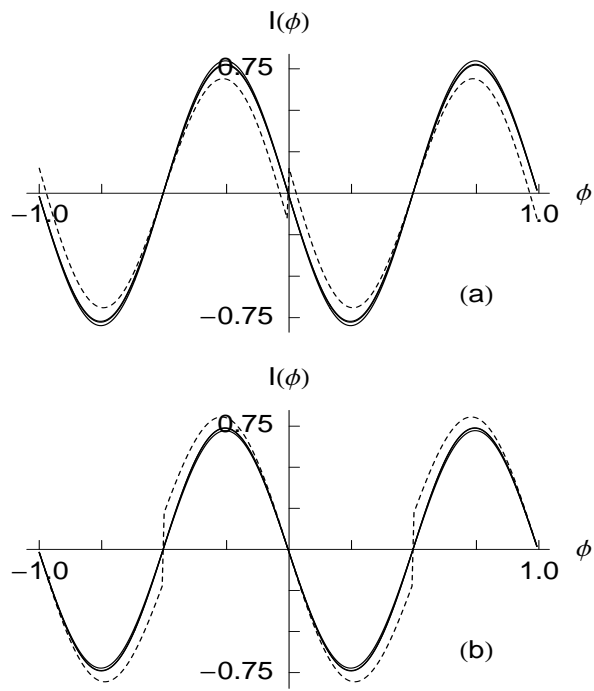}}\par}
\caption{ $I-\phi$ curves for the same systems as those 
in Fig.~\ref{random}
excepting that the $\epsilon_i$'s are chosen randomly from Eq.~(\ref{site2}).}
\label{binary}
\end{figure}
 Now we address the problem of persistent current in SID mesoscopic rings 
using the Hamiltonian Eq.~(\ref{hamil}), and, in this study we do not 
consider the spin of the electrons as it will not change the qualitative 
behavior of the currents within the one-electron picture. We present exact 
calculation of the currents in the presence of all LRH integrals in the
tight-binding Hamiltonian with diagonal disorder, and it involves exact 
numerical diagonalization of the Hamiltonian matrices. The results for
some representative examples are given in Fig.~\ref{random} and 
Fig.~\ref{binary}. In Fig.~\ref{random}, we plot $I(\phi)$ versus $\phi$ 
curves for the systems with $v=-1,~ \mu=1.4, ~N=50$, and, $N_e=20$ and $23$, 
where disorder is introduced by random choice of $\epsilon_i$'s from the 
``box" distribution Eq.~(\ref{site1}) setting $W=1$. The solid lines 
correspond to three microscopic configurations of disorder, while the dotted 
lines are for the ordered cases obtained by setting all $\epsilon_i$'s equal 
to zero. Fig.~\ref{binary} is the $I-\phi$ characteristics of the systems 
with the same set of parameters as those in Fig.~\ref{random}, where site 
energies are chosen randomly from the ``binary-alloy" distribution
given by Eq.~(\ref{site2}) with $c=0.5$ and $\delta=|\epsilon_A-\epsilon_B|=1$.
We consider three typical disordered configurations of the ring compatible
with the ``binary-alloy" distribution, and the $I-\phi$ curves for these
configurations are represented by solid lines in Fig.~\ref{binary}. 
The dotted lines in this figure are identical to those in Fig.~\ref{random}. 

 Let us now analyze the results presented in Fig.~\ref{random} and
Fig.~\ref{binary}. We see that for the present model of SID mesoscopic ring, 
the persistent currents are always periodic in $\phi$ with periodicity 
$\phi_0$. Fig.~\ref{random} and Fig.~\ref{binary} clearly show that in the 
presence of all LRH integrals in the Hamiltonian, the $I-\phi$ characteristics 
of a given SID mesoscopic ring are almost insensitive to the microscopic 
configuration of disorder of the ring, and, we have checked considering 
$100$ distinct configurations of the given system that all the $I-\phi$ 
curves nearly collapse to a single curve. These figures 
are self-explanatory to revel the fact that as we vary the microscopic
configurations of the ring, the persistent currents do not fluctuate in 
sign and the fluctuation in magnitude becomes exceedingly small. The most 
interesting result is that in the present model persistent current is not 
reduced by disorder, and it is apparent from Fig.~\ref{random} and 
Fig.~\ref{binary} that the currents in the disordered rings are of the 
same order of magnitude as the current in the ordered ring. We have also 
seen that the decrease in amplitude of the current is quite small even 
if we increase the strength of disorder. These results dramatically differ 
from the previous predictions according to which the persistent currents 
exhibit strong fluctuations both in sign and magnitude depending on the 
realization of disorder, and the currents are reduced by several orders of 
magnitude due to disorder. The orders of magnitude reduction of the persistent 
currents suggested earlier are basically due to the tendency of localization
of the electrons in the Anderson-like models. On the other hand, the
present tight-binding model with all LRH integrals supports extended
electronic eigenstates even in the presence of disorder~\cite{bala}, 
and for this reason persistent currents are not reduced by the impurities. 
Our results
are in good agreement with the experimental observations~\cite{chand,maily}. 

 We have also noticed certain interesting features of the $I-\phi$ curves 
that are characteristics of any disordered ring. The discontinuity in 
$I(\phi)$ as a function of $\phi$ are characteristics of the ordered 
systems which disappears due to disorder, and the current in the
disordered rings are always zero at these points of discontinuity. This
result can be easily understood on a very general ground. We can treat
disorder as a perturbation over the ordered situation that lifts the
degeneracy at the crossing points of the unperturbed energy levels. So
gaps open up at the ground level in the presence of disorder making
$I(\phi)$ a continuous function of $\phi$, and, also $I(\phi)$ becomes 
exactly equal to zero at these points of discontinuity.

 In conclusion, we have investigated the behavior of persistent currents
in SID mesoscopic rings by a simple model including all LRH integrals
in the usual Anderson model. Our exact calculations show that both the
sign and magnitude of the experimentally observed currents can be explained
from the present model. In this work we have convincingly established that 
the essential physical mechanisms are the LRH integrals that accounts for 
the observed behavior of persistent currents in SID mesoscopic rings.

\end{document}